\documentclass[article]{IEEEtran}

\usepackage{cite}

\usepackage{tikz} 
\usepackage{pgfplots}
\usepackage{tumcolor}

\usepackage[cmex10]{amsmath}
\usepackage{amssymb}
\usepackage{fixmath}
\usepackage{amsbsy}
\usepackage{algorithmicx}
\usepackage{algpseudocode}
\usepackage{algorithm}
\usepackage{booktabs}
\usepackage{bm}

\usepackage{array}

\ifCLASSOPTIONcompsoc
  \usepackage[caption=false,font=normalsize,labelfont=sf,textfont=sf]{subfig}
\else
  \usepackage[caption=false,font=footnotesize]{subfig}
\fi

\usepackage{dane}

\pgfplotsset{compat=1.13}

\begin{document}

\title{Covariance Matrix Estimation in Massive MIMO}

\author{David Neumann, Michael Joham, and Wolfgang Utschick\\
Methods of Signal Processing,
	    Technische Universit\"at M\"unchen,
	      80290 Munich, Germany\\\small
	        \{\texttt{d.neumann,joham,utschick}\}\texttt{@tum.de}
	}

\markboth{}%
{}

\maketitle

\begin{abstract}
    Interference during the uplink training phase significantly deteriorates the performance of a massive MIMO system.
    The impact of the interference can be reduced by exploiting second order statistics of the channel vectors, e.g., to obtain minimum mean squared error estimates of the channel.
    In practice, the channel covariance matrices have to be estimated.
    The estimation of the covariance matrices is also impeded by the interference during the training phase.
    However, the coherence interval of the covariance matrices is larger than that of the channel vectors.
    This allows us to derive methods for accurate covariance matrix estimation by appropriate assignment of pilot sequences to users in consecutive channel coherence intervals.
\end{abstract}

\IEEEpeerreviewmaketitle

\section{Introduction} 

Massive MIMO is currently drawing a lot of interest from both, academia and industry, due to the promise of full multiplexing gains with simple linear signal processing methods~\cite{rusek_scaling_2013}.
This gain, it seems, is limited by the dimensionality bottleneck imposed by the fixed coherence interval of the channel~\cite{zheng_communication_2002}.
However, the well-known results on the degrees of freedom achievable in a fixed coherence interval only consider identically distributed channel coefficients.
In fact, structure of the channel vectors in form of second-order information can be exploited to break out of the dimensionality bottleneck~\cite{huh_achieving_2012,neumann_mse_2017,neumann_ratebalancing_2015}.

Due to the large dimensionality of the channels, the covariance matrix estimation might be difficult to realize in practice.
Furthermore, to estimate the covariance matrix of a single channel vector to one user, we also need to deal with the contamination in the observations.
For specific channel models and communication scenarios, such as wide-band scenario with sparse channels~\cite{haghighatshoar_massive_2017-1} it might be possible to deal with the interference as long as certain separability conditions hold.
Approaches which rely on additional pilot transmission offer a way to estimate the covariance matrices which does not depend on the properties of the channel~\cite{shariati_low-complexity_2014-1,bjornson_massive_2016-1,neumann_lowcomplexity_2015}.

We propose a method for covariance matrix estimation that is able to deal with contaminated observations without additional pilot overhead and without strong assumptions on the channel model.
The estimation method is based on a systematic allocation of pilot sequences to users in consecutive coherence intervals.
We first show that a two-step estimation procedure, where we first estimate the covariance matrices of the contaminated observations and then the covariance matrices of the channel vectors, allows us to accurately identify the channel covariance matrices.
Then we derive the optimality conditions of the maximum-likelihood estimator and propose a fixed-point approach to solve the resulting non-linear system of equations.
We also introduce an adaptive algorithm for practical application.

The effectiveness of the proposed estimation methods is demonstrated with numerical simulations.
We show that, with our approach to covariance matrix estimation, we can achieve performance close to that of perfect knowledge of the second order statistics.

\section{System Model}

A base station with $M$ antennas wants to estimate the covariance matrices (and channel vectors) of $K$ single antenna users.
Those $K$ users can include strong interferers in neighboring cells, which are not served by the base station, but have significant influence on the achievable performance.
The system operates in time-division duplex mode which allows to exploit the reciprocity of the channel.
In each coherence interval, the users transmit pilot sequences during an uplink training phase.
We have $\Ttr$ orthonormal pilot sequences available and assume that $\Ttr < K$.
Consequently, at least two users share the same pilot sequence.
In a multi-cell scenario, the allocation of pilot sequences in neighboring cells has to be known.
This can be achieved by a deterministic network-wide schedule or by exchange of information over a backhaul link.

Let $\vPi[t] \in \{0,1\}^{K \times \Ttr}$ denote the allocation of pilot sequences to users in coherence interval $t$.
That is, $[\vPi[t]]_{kp} = 1$ indicates that user $k$ transmits the pilot sequence $p$ in coherence interval $t$.
Since each user transmits one of the pilot sequences in each channel coherence interval, each row of $\vPi[t]$ has exactly one non-zero entry.

In coherence interval $t$, we correlate the received training signals with the pilot sequences to obtain the observations
\begin{equation}\label{eq:system_model}
    \vPhi[t] = \big[\vobs_1[t],\ldots,\vobs_{\Ttr}[t] \big] = \vH[t] \vPi[t] + \vV[t] \in \mathbb C^{M\times \Ttr}.
\end{equation}
The compound channel matrix
    $\vH[t] = \big[ \vh_1[t],\ldots, \vh_K[t] \big] \in \mathbb C^{M\times K}$
is composed of the channel vectors of all $K$ users  in coherence interval $t$ and $\vV[t]$ is additive white Gaussian noise with i.i.d. zero-mean entries with variance $\sigma_v^2$.
The channel vectors of different users and different coherence intervals are assumed to be independently distributed with $\vh_k[t] \sim \CN(\zeros, \vC_{\vh_k})$.

\section{Problem Formulation}

Since $\Ttr < K$, i.e., the $\vPi[t]$ are tall matrices, we suffer from interference in the observations $\vPhi[t]$.
It is thus desirable to use second-order statistics to estimate the channel vectors and design the transmit and receive filters.
The channel covariance matrices can be used to calculate the minimum mean squared error (MMSE) estimate of the channel vector $\vh_k[t]$ of user $k$ based on the observations $\vPhi[t]$, namely,
\begin{equation}\label{eq:mmse_est}
    \hat \vh_k[t] = \vC_{\vh_k} \vC_{\vobs_p[t]}\inv \vobs_p[t] \in \mathbb C^M.
\end{equation}
where $p$ is the index of the pilot sequence transmitted by user $k$, i.e., $[\vPi[t]]_{kp} = 1$, and $\vC_{\vobs_p[t]} = \expec[\vobs_p[t]\vobs_p[t]\he]$.
However, in practice, the covariance matrices are unknown.

Estimating the covariance matrices is a challenging problem due to the large number of parameters that have to be estimated and due to the interference during the training phase.
We reduce the number of parameters by assuming diagonal covariance matrices.
This assumption is reasonable for certain array geometries that allow to jointly diagonalize the covariance matrices of all users for large numbers of antennas.
For example, for a uniform linear array under the far-field assumption, all channel covariance matrices have Toeplitz structure and are approximately diagonalized by the discrete Fourier transform~\cite{neumann_lowcomplexity_2015}.
Thus, if we perform all signal processing in the frequency domain of the array, we can approximate the covariance matrices by diagonal matrices with reasonable accuracy.
Also, for distributed antennas, we typically have diagonal covariance matrices~\cite{ngo_cellfree_2015}.

In the following, all covariance matrices are of the form
%\begin{equation*}
    $\vC_\vx = \diag(\vc_\vx) \in \mathbb C^{M\times M}$
%\end{equation*}
with $\vc_\vx \in \mathbb C^M$.
Note that in principle, the proposed methods also work for full covariance matrices, but just working with the diagonals simplifies all aspects of estimation and transeiver design significantly.
Recent work~\cite{bjornson_massive_2017} demonstrates that ignoring off-diagonal elemnts still leads to asymptotically optimal achievable rates.

In the massive MIMO literature, it is typically assumed that a fixed allocation $\vPi[t] = \vPi$ is reused in every coherence interval. 
This is reasonable for constant and known channel covariance matrices, 
since in this case, an optimal assignment with respect to the considered network utility function can be found.

If the same assignment is reused in every time slot, the columns $\vobs_p[t]$ of $\vPhi[t]$ are identically distributed for each $t$.
Thus, we can estimate the corresponding covariance matrices $\vC_{\vobs_p}=\diag(\vc_{\vobs_p})$ by straightforward maximum-likelihood (ML) estimation.

Let us collect the diagonals of all channel covariance matrices into the matrix
\begin{equation}
    \vC = [\vc_{\vh_1},\ldots,\vc_{\vh_K}] \in \mathbb C^{M\times K}.
\end{equation}
For known noise variance $\sigma_v^2$, we have the relation [cf.~\eqref{eq:system_model}]
\begin{equation}\label{eq:coveq_same_alloc}
    [\vc_{\vobs_1}, \ldots, \vc_{\vobs_{\Ttr}}] - \sigma_v^2\ones = \vC \vPi
\end{equation}
where $\ones$ is the $M\times \Ttr$ all-ones matrix.
Due to $\vPi$ being a tall matrix, we fail to reconstruct unique channel covariance matrix estimates  from the estimated covariance matrices of the observations.
This indicates that using the same allocation $\vPi$ in each time-slot leads to an ill-posed problem for the covariance matrix estimation.

\section{Two-step Covariance Matrix Reconstruction}
\label{sec:reconstruction}

Now suppose we iterate through a set of different allocations $\vPi_i$, with $i = 1,$ \dots, $N$.
This results in blocks of $N\Ttr$ observations $\vobs_\tau$, $\tau=1,\ldots,N\Ttr$, each with a different covariance matrix.
The relation between the covariance matrices expands to~[cf.~\eqref{eq:coveq_same_alloc}]
\begin{equation*}
    [\vc_{\vobs_1}, \ldots, \vc_{\vobs_{N\Ttr}}] - \sigma_v^2\ones = \vC [\vPi_1, \ldots, \vPi_N].
\end{equation*}
As long as the joint allocation matrix $[ \vPi_1, \ldots, \vPi_N]$ has full row rank, we can apply a right inverse to reconstruct the channel covariance matrices
\begin{equation}\label{eq:two_step_solution}
     \vC  = \big( [ \vc_{\vobs_1}, \ldots,  \vc_{\vobs_{N\Ttr}}] - \sigma_v^2 \ones \big) [\vPi_1,\ldots,\vPi_N]^+.
\end{equation}
In other words, with a proper choice of the pilot allocations, the reconstruction of the channel covariance matrices can be turned into a well-conditioned problem.

\bigskip
\noindent \textit{Example}. Suppose we have $\Ttr=2$ training sequences and $K=4$ users.
This setup allows $N=3$ distinct allocations of pilots to users
\begin{equation*}
    \vPi_1 = \mat{
        1 & 0 \\
        1 & 0 \\
        0 & 1 \\
        0 & 1
    }, \quad
    \vPi_2 = \mat{
        1 & 0 \\
        0 & 1 \\
        1 & 0 \\
        0 & 1
    }, \quad
    \vPi_3 = \mat{
        1 & 0 \\
        0 & 1 \\
        0 & 1 \\
        1 & 0
    }
\end{equation*}
The compound matrix $[\vPi_1,\vPi_2,\vPi_3]$ is well-conditioned with a condition number of $\sqrt{3}$.

\section{ML Problem}

In this section, we consider the general maximum-likelihood estimation of the channel covariance matrices for time-varying pilot assignments $\vPi[t]$.
We assume a minimum coherence interval of the covariance matrices of $T$ channel coherence intervals, i.e., we use the observations of the last $T$ channel coherence intervals to estimate the covariance matrices and assume that the covariance matrices are constant for this time interval.
In the general case, we have different allocations $\vPi[t]$ and resulting observations $\vPhi[t]$ for $t=1,$ \dots, $T$.
For notational convenience, we collect the allocations and observations in the matrices
\begin{align}
    \vPi &= [\vPi[1],\ldots, \vPi[T]] = [\vpi_1, \ldots, \vpi_{\Ttr T}] \in \mathbb C^{K \times T\Ttr}\\
    \vPhi &= [\vPhi[1],\ldots, \vPhi[T]] = [\vobs_1, \ldots, \vobs_{\Ttr T}] \in \mathbb C^{M \times T\Ttr}
\end{align}
respectively.
Note that the observations $\vobs_i$ are mutually independent, due to the block fading model for the channel vectors and the fact that within one channel coherence interval each user only transmits one pilot sequence.
Additionally, since we assume diagonal covariance matrices, the rows of $\vPhi$ are mutually independent.

As a consequence, the maximum likelihood problem to estimate the diagonal covariance matrices $\vC$ [see~\eqref{eq:coveq_same_alloc}] can be separately solved for each row of $\vC$.
With $\vC = [\vc_1, \ldots, \vc_M]\tp$ and $\vPhi = [\bar \vobs_1, \ldots, \bar \vobs_M]\tp$, we get
\begin{equation}
    \min_{\vC} L(\vC; \vPhi) = \min_{ \vC} \sum_m L_m(\vc_m; \bar \vobs_m) = \sum_m \min_{\vc_m} L_m(\vc_m; \bar \vobs_m).
\end{equation}
With the vector of element-wise absolute squared observations $\vb_m = \abs{\bar \vobs_m}^2$, the $m$-th negative log-likelihood function (LLF) is 
\begin{equation}\label{eq:likelihood}
    L_m(\vc_m; \vb_m) = \sum_{i=1}^{T\Ttr} \left( \frac{[\vb_m]_i}{ \vc_m\tp \vpi_i + \sigma_v^2} + \log (\vc_m\tp \vpi_i + \sigma_v^2 ) \right).
\end{equation}
Setting the derivative of the LLF to zero yields 
\begin{align}
    \label{eq:deriv_ml}
    \frac{\partial L(\vc_m;\vb_m)}{\partial \vc_m} = \sum_i \frac{\vc_m\tp\vpi_i + \sigma_v^2 - [\vb_m]_i}{ (\vc_m\tp \vpi_i + \sigma_v^2)^2} \vpi_i = \zeros
\end{align}
which we reformulate to
\begin{equation}\label{eq:sys}
    \frac{1}{T} \left(\sum_i \vpi_i d_{m,i} \vpi_i\tp\right) \vc_m = \frac{1}{T}\sum_i \vpi_i d_{m,i} ([\vb_m]_i - \sigma_v^2) 
\end{equation}
or in matrix-vector notation
\begin{equation}
    \frac{1}{T}\vPi \vD_m \vPi\tp \vc_m = \frac{1}{T}\vPi \vD_m (\vb_m - \sigma_v^2 \ones)
\end{equation}
with $\vD_m = \diag(d_{m,1},\ldots,d_{m,T\Ttr})$ and
\begin{equation*}
    d_{m,i} = \frac{1}{(\vpi_i\tp \vc_m + \sigma_v^2)^2}.
\end{equation*}
The apparent issue is that the desired variances $\vc_m$ appear in the denominator of the $d_{m,i}$. 
Thus, finding a solution to the non-linear system of equations in~\eqref{eq:sys} seems difficult.
In fact, the cost function in~\eqref{eq:likelihood} is neither convex nor quasi-convex. 
Therefore, we have no guarantee of a unique global optimum.

However, since the $[\vb_m]_i$ are independent and have bounded variance, the strong law of large numbers applies for $T \rightarrow \infty$, i.e.,
\begin{equation*}
    \frac{1}{T} \vPi \vD_m \vb_m - \frac{1}{T}\expec[\vPi \vD_m \vb_m] \xrightarrow{\text{a.s.}} \zeros.
\end{equation*}
With
\begin{equation*}
    \frac{1}{T}\expec[\vPi \vD_m \vb_m] = \frac{1}{T}\vPi \vD_m (\vPi\tp \vc_m + \sigma_v^2 \ones)
\end{equation*}
we get 
\begin{equation*}
    \hat \vc_m = (\frac{1}{T}\vPi \vD_m \vPi\tp)\inv \frac{1}{T}\vPi \vD_m (\vb_m - \sigma_v^2 \ones) \xrightarrow{\text{a.s.}} \vc_m
\end{equation*}
as long as $\abs{d_{m,i}}^2 < \infty, \forall i$ and $\frac{1}{T}\vPi \vD_m \vPi\tp$ is invertible in the limit.
In other words, given a large number of observations, the estimate $\hat \vc_m$ converges to the true variances $\vc_m$ irrespective of the scaling matrix $\vD_m$.
However, if only a few observations are available, we might get better estimates by choosing a scaling matrix similar to the $\vD_m$ suggested by the ML optimality condition.

In practice, the estimate $\hat \vc_m$ is updated in each channel coherence interval.
This suggests an adaptive algorithm, that updates the averages on the left and right-hand-side of~\eqref{eq:sys} in each channel coherence interval and which always uses the estimate from the last coherence interval to determine the scaling coefficients $d_{m,i}$.
The algorithm is described in detail in Alg.~\ref{alg:adaptive}. 
Note that in a practical implementation, we would not update the matrix $\vXi = \vPi \vD\vPi\tp$ but instead directly update its Cholesky decomposition.
This reduces computational complexity and increases numerical stability.

\begin{algorithm}
    \caption{Adaptive Variance Estimation}
    \label{alg:adaptive}
    \begin{algorithmic}[1]
        \State Similarly to many adaptive algorithms we use a constant forgetting factor $0 < \lambda < 1$.
        \State Initialize the estimate of the matrix $\vXi = \vPi \vD \vPi\tp$ 
        \[
            \vXi \assign \id
        \]
        \State Initialize the variance estimates and accumulated observations
        \[
            \hat \vc_m \assign \ones \quad \vpsi \assign \zeros
        \]
        \For{$t=1,\ldots$}
            \State Acquire the current allocation $\vPi[t]$ and resulting observations $\vb_m[t]$ at antenna $m$
            \State Calculate the approximate scaling matrix using the current variance estimates
            \[
                [\vd]_p \assign 1/(\vpi_p[t]\tp \hat\vc_m + \sigma_v^2)^2, \forall p=1,\ldots,\Ttr
            \]
            \State Update the accumulated observations
            \[
                \vpsi \assign \lambda \vpsi + \vPi[t] \diag(\vd) (\vb_m[t] - \sigma_v^2 \ones)
            \]
            \State Update the matrix $\vXi$
            \begin{align*}
                \vXi &\assign \lambda\vXi + \vPi[t] \diag(\vd) \vPi[t]\tp
            \end{align*}
            \State Calculate the new variance estimates
            \[
                    \hat \vc_m \assign \vXi\inv \vpsi
                \]
        \EndFor
    \end{algorithmic}
\end{algorithm}

By choosing a different scaling matrix for each dimension, we might get better results, but suffer in terms of memory overhead and computational complexity.
Thus, for practical purposes, we might choose the same scaling matrix $\vD_m = \vD$ for each dimension.
The estimation of all desired parameters then simplifies to
\begin{equation} \label{eq:fullest}
    \hat \vC = (\vB - \sigma_v^2\ones ) \vD \vPi\tp (\vPi \vD \vPi\tp)\inv 
\end{equation}
where $\vB = [\vb_1, \ldots,\vb_M]\tp$.

If the same sequence $\widetilde \vPi = [\vPi_1, \ldots,\vPi_N]$ of allocations [cf.~Section~\ref{sec:reconstruction}] is repeated in $S$ blocks, such that $SN = T$, we
group the observations into equally sized blocks $\vB = [ \vB_1, \ldots, \vB_S]$ to get 
\begin{equation*}
    \hat \vC = \Big( \frac{\sum_{s=1}^S \vB_s}{S} - \sigma_v^2\ones \Big) \widetilde \vD \widetilde \vPi\tp (\widetilde \vPi \widetilde \vD \widetilde \vPi\tp)\inv.
\end{equation*}
Since 
\begin{equation*}
    \frac{\sum_{s=1}^S \vB_s}{S} = [\hat\vc_{\vobs_1},\ldots,\hat\vc_{\vobs_{N\Ttr}}]
\end{equation*}
is exactly the ML estimate of the covariance matrices of the observations,
the suboptimal estimation in~\eqref{eq:fullest} has the two-step reconstruction method from Section~\ref{sec:reconstruction} as special case, namely when we chose $\widetilde\vD = \id$ [see~\eqref{eq:two_step_solution}].

\section{Pilot Allocation}

If we implement the covariance matrix estimation with a repeated schedule $\widetilde \vPi = [\vPi_1, \ldots, \vPi_N]$ of pilot allocations [cf.~Section~\ref{sec:reconstruction}],
$\widetilde \vPi$ must have full row-rank for unique identifiability of the channel covariance matrices.
If we want to serve all $K$ users within one coherence interval, each user has to be assigned a pilot sequence, i.e., $\vPi_t \ones = \ones$ for all $t$. 
Consequently, by adding one coherence interval to the schedule $\widetilde \vPi$, the rank increases at most by $\Ttr-1$.
We have 
\begin{equation}\label{eq:cond_N}
    \rank(\widetilde\vPi) \leq \Ttr + (N-1)(\Ttr-1)
\end{equation}
and we need $\rank(\widetilde \vPi) = K$, leading to the necessary condition
\begin{equation}\label{eq:bound_on_N}
    N \geq \frac{K-1}{\Ttr-1}.
\end{equation}
Thus, when we serve all users in each time-slot, we need at least two training sequences to ensure full row-rank of $\widetilde \vPi$.

Since the schedules can be calculated off-line for given $K$, $\Ttr$, and the resulting schedule interval $N$, we could theoretically do an exhaustive search over all feasible schedules to find the schedule with best condition number of $\widetilde \vPi$.
However, the design of the allocation is not actually an issue in practice.
Typically the coherence interval of the covariance matrices is much larger than the bound in~\eqref{eq:bound_on_N}.
Thus, if we use a slightly larger $N$ and random allocations, we get a full-rank matrix $\widetilde \vPi$ with high probability.

\section{Results}

To compare with the previously introduced approach in~\cite{bjornson_massive_2016-1} we perform simulations in the same multi-cell setup with seven hexagonal cells.
We use the same setup of $K_C=10$ users per cell in a ring around the base station which employs a uniform linear array with $M=100$ antennas.
We have thus $K=70$ users in the local neighborhood of the network of which we want to estimate the covariance matrices.

As performance metric we calculate the achievable sum-rate in the uplink (cf.~\cite{bjornson_massive_2016-1}) in the center cell for a regularized zero-forcing (RZF) receive filter.
The filter is calculated based on the available channel estimates.
As one base-line, we use the simple least-square (LS) estimation that does not require covariance matrix information.
For all other methods, the (approximate) MMSE estimates of the channel vectors are used, where the channel estimates are calculated using the available covariance matrix estimates.
We also include a genie-aided approach, which uses the actual covariance matrices.

\setlength{\textfloatsep}{2mm}
\setlength{\abovecaptionskip}{0pt}
\begin{figure}[t]
    \begin{center}
        \begin{tikzpicture}
\begin{axis}[
    height=0.70\columnwidth,
    width=1.00\columnwidth,
    grid=both,
    ticks=both,
    xmin=20,
    xmax=210,
    ylabel={\small Achievable sum-rate},
    xlabel={\small Covariance matrix coherence interval $T$},
    title={},
    yticklabel style={/pgf/number format/.cd, fixed, fixed zerofill, precision=0},
    legend entries={
        \footnotesize Genie-aided,
        \footnotesize Approximate ML estimation,
        \footnotesize Two-step reconstruction,
        \footnotesize Method from~\cite{bjornson_massive_2016-1},
        \footnotesize LS channel estimation,
    },
    legend style={at={(1.00,0.00)}, anchor=south east},
]

\addplot[ color=TUMBeamerRed, dotted, line width = 1.5pt ]
    table [x index=0,y index=5,col sep=comma] {multi_wrt_T.csv};
\addplot[ color=TUMBeamerOrange, line width = 1.5pt, mark=* ]
    table [x index=0,y index=3,col sep=comma] {multi_wrt_T.csv};
\addplot[ color=TUMDarkerBlue, line width = 1.5pt, mark=triangle ]
    table [x index=0,y index=2,col sep=comma] {multi_wrt_T.csv};
\addplot[ color=black, dash pattern=on 5pt off 1pt on 1pt off 1pt, line width=1.5pt ]
    table [x index=0,y index=7,col sep=comma] {multi_wrt_T.csv};
\addplot[ color=black, dotted, line width=1.5pt ]
    table [x index=0,y index=1,col sep=comma] {multi_wrt_T.csv};

\end{axis}
\end{tikzpicture}
    \caption{
        Achievable sum-rate with respect to the number of channel coherence intervals used to estimate the covariance matrices. 
        We use the given multi-cell scenario with $\Ttr = 11$ pilot sequences per training phase.
    }
    \label{fig:multi_wrt_T}
\end{center}
\end{figure}
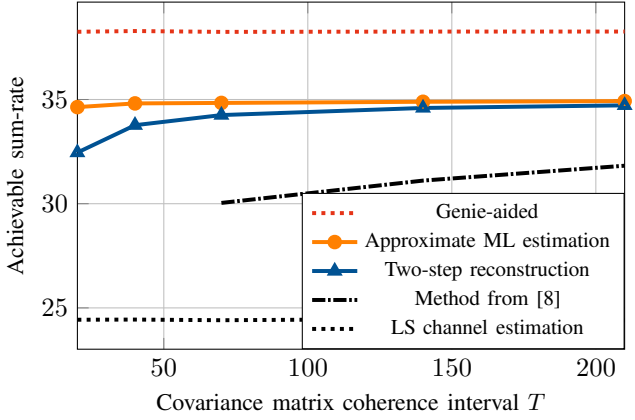

Our simulations show that a relatively small coherence interval for the covariance matrices is sufficient to get accurate covariance matrix estimates.
In Fig.~\ref{fig:multi_wrt_T}, we illustrate the performance with respect to the covariance matrix coherence interval $T$.
That is, we show the results with respect to the number of training phases which are used to estimate the covariance matrices.
In each training phase, we use $\Ttr = K_C+1$ orthogonal pilot sequences for training.
For the methods from~\cite{bjornson_massive_2016-1}, the extra pilot sequence is used to estimate the covariance matrices.
Thus, if the covariance matrices are quasi-constant for $T=Kn$ channel coherence intervals, we get $n$ observations per user to facilitate the covariance matrix estimation.
For $T < K$, the methods from~\cite{bjornson_massive_2017} cannot produce covariance estimates for all users.
In our approach there is no distinction between pilots used for channel estimation and pilots used for covariance matrix estimation.
For each training phase, we simply allocate pilots randomly to users such that users in the same cell use different pilots.

The performance of all covariance matrix estimation methods is significantly higher than that in~\cite{bjornson_massive_2016-1} because we use the circulant approximation for all methods instead of the regularization proposed in~\cite{bjornson_massive_2016-1}. 
While the method from~\cite{bjornson_massive_2016-1} only yields results for $T\geq K=70$, the approximate ML approach already achieves peak performance for much smaller coherence intervals.
There is a small constant gap to the genie-aided approach due to the circulant approximation.
This is a small price to pay for improved estimation accuracy for a small number of observations and significantly lower computational complexity.

\begin{figure}[t]
    \begin{center}
        \begin{tikzpicture}
\begin{axis}[
    height=0.70\columnwidth,
    width=1.00\columnwidth,
    grid=both,
    ticks=both,
    xmin=10,
    xmax=14,
    ylabel={\small Achievable sum-rate },
    xlabel={\small Number of training sequences $\Ttr$}, 
    title={},
    yticklabel style={/pgf/number format/.cd, fixed, fixed zerofill, precision=0},
    legend entries={
        \footnotesize Perfect CDI,
        \footnotesize Approximate ML estimation,
        \footnotesize Two-step reconstruction,
        \footnotesize Method from~\cite{bjornson_massive_2016-1},
        \footnotesize LS estimation,
    },
    legend style={at={(1.00,0.00)}, anchor=south east},
]

\addplot[ color=TUMBeamerRed, dotted, line width = 1.5pt ]
    table [x index=0,y index=5,col sep=comma] {multi_wrt_Ttr.csv};
\addplot[ color=TUMBeamerOrange, line width = 1.5pt, mark=* ]
    table [x index=0,y index=3,col sep=comma] {multi_wrt_Ttr.csv};
\addplot[ color=TUMDarkerBlue, line width = 1.5pt, mark=triangle ]
    table [x index=0,y index=2,col sep=comma] {multi_wrt_Ttr.csv};
\addplot[ color=black, dash pattern=on 5pt off 1pt on 1pt off 1pt, line width=1.5pt ]
    table [x index=0,y index=7,col sep=comma] {multi_wrt_Ttr.csv};
\addplot[ color=black, dotted, line width=1.5pt ]
    table [x index=0,y index=1,col sep=comma] {multi_wrt_Ttr.csv};

\end{axis}
\end{tikzpicture}
    \caption{
        Achievable sum-rate with respect to the number of orthogonal training sequences $\Ttr$ per channel coherence interval.
        The covariance matrix coherence interval is fixed to $T=70$ channel coherence intervals.
    }
    \label{fig:multi_wrt_Ttr}
\end{center}
\end{figure}
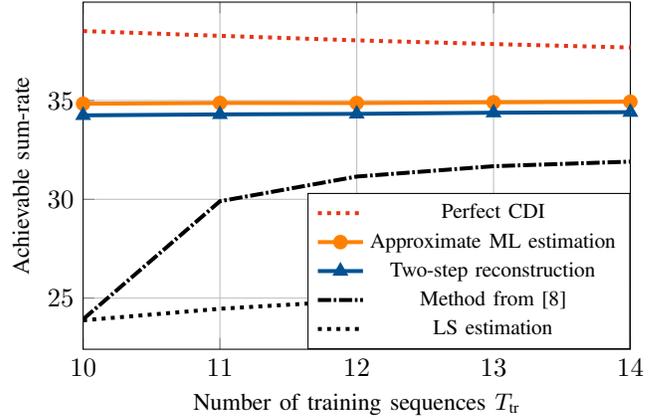

In Fig.~\ref{fig:multi_wrt_Ttr}, we show the performance with respect to the number of orthogonal pilot sequences $\Ttr$ per training phase for a fixed coherence interval of $T=70$.
We start with $\Ttr = K_C$, for which the approach from~\cite{bjornson_massive_2016-1} does not give a satisfactory result, since there are no extra pilots that can be used for covariance matrix estimation.
Our approach, however, is able to find reasonably accurate estimates of the covariance matrices even when the number of available pilot sequences is exactly the number of users per cell.
For the genie aided approach, the achievable rate decreases with $\Ttr$, due to the reduced number of channel accesses available for data transmission.

\section{Conclusion}

We demonstrated that the newly introduced methods for covariance estimation allow to accurately estimate channel covariance matrices from a small number of observations, even with interference during the training phase.
As a consequence, the proposed methods enable approaches that need accurate second order statistics to deal with pilot-contamination in massive MIMO systems.

\bibliographystyle{IEEEtran}
\bibliography{IEEEabrv,literature}

\end{document}